# Visualizable Detection of Nanoscale Perturbations

Jinlong Zhu[1,*,†], Aditi Udupa[1], Lynford L. Goddard[1,*]

[1]Photonic Systems Laboratory, Micro and Nanotechnology Laboratory, Department of Electrical and Computer Engineering, University of Illinois at Urbana-Champaign, Urbana, Illinois 61801, USA

[*]These authors contributed equally to this work.

Corresponding author: [†]zhuwdwz1@illinois.edu

**Abstract:** Using light to non-destructively detect nanoscale perturbations is vital to many fields including material characterization, human disease diagnosis, and semiconductor electronics. In this work, we introduce the concepts of electromagnetic canyons and non-resonance amplification and apply them on a conventional diffraction-limited optical microscope to directly view individual perturbations (25-nm radius = $\lambda/31$) in a nanoscale volume. Considering the extensive impact of microscopy on scientific discovery and technology development, our noninvasive imaging-based method with deep subwavelength footprint will have far-reaching consequences that will affect our everyday lives.

**Main Text:** A perturbation is a small change in the steady state of a physical system. Fixed-form nanoscale objects such as semiconductor defects, environmental dust particles, and voids in integrated circuits *(1)*, and free-form objects such as nanoparticles, viruses, and molecule clusters in water and tissue fluids *(2)*, can be treated as perturbations because they result in an alteration on the function of the host media. The detection for these perturbations provides a feedback mechanism to control the fabrication quality of functional devices as well as to understand the physical, biological, or chemical roles acted by perturbations; thus, it is extremely vital to many fields including semiconductor circuits, integrated photonics, quantum chips, human disease diagnosis, medicine, and security systems, to name a few. Moreover, there is an increasing interest from industry and academia of not only detecting perturbations, but also visualizing them in nanometric volumes. This enables the non-contact classification for perturbations as well as the monitoring of sensing dynamics at a single object level. Typical examples include the visualizable classification of deep subwavelength defects in semiconductor industry *(3)* and virus/supermolecule counting *(4)*. However, such a stringent requirement hinders the application of the well-established non-visualizable spectrum-alteration-based resonance sensors *(5-12)*. Electron microscopy offers nanometric resolution, but it is destructive, slow, costly, and has an ultra-small field-of-view for nanoscale perturbations. Optical super-resolution imaging techniques are intrinsically fast, but even the state-of-the-art fluorescence microscopy *(13-15)* and scanning nearfield optical microscopy may not fit because of the phototoxicity, instrument complexity, secondary pollution, and incompatibility to massive nanofabrication and sensing *(3)*. Photothermal imaging has been demonstrated for label-free imaging of individual particles *(16, 17)*, but the accumulated Joule heat by a high-Q cavity may cause damage to biomaterials and even functional background media. In summary, we need a brand-new visualizable detection and sensing modality that meets nearly all the stringent requirements, i.e., label-free, fast, non-destructive, Joule-heat-free, simple in instrumentation, cheap, easy-to-operate, convenient-to-integrate, and large in field-of-view.

In this paper, we propose such a visualizable modality, which uses only two simple add-on apparatuses, in order to revolutionize conventional low-performance optical microscopes and enable detecting nanoscale perturbations in nanometric volumes. Conventionally, people believe that using diffraction-limited optical microscopes to detect nanometric perturbations is extremely challenging because of not only the diffraction barrier, but also the weak Rayleigh scattering *(1)*, i.e., a $d^6/\lambda^4$ scaling of the detectable far-field signal is inevitable for a particle with size *d*. This is true, because the extremely weak scattering signal (for instance, the intensity, as vividly represented by a crab in Fig. 1A) can be easily overwhelmed by the background signal that is induced by the scattering from surrounding patterns/substrates as well as by fluctuations induced by system errors and the instability of the instruments *(3)*. See the artwork in Fig. 1A. Our proposal, however, breaks these barriers by artificially creating an "electromagnetic canyon" (EC; the region where background electromagnetic field is null) using a two-beam illumination apparatus, such that the far-field scattering of a nanoscale perturbation is amplified by a non-resonance nanostructure ensemble (behaves as the second apparatus) and thus can be directly imaged in the conventional optical microscope. See the artwork in Fig. 1B showing the concept of ECs. The non-resonance amplification is different than the principle of resonance sensors [such as ring resonators and metallic particles *(5-10)*, which work by tuning the spectrum around resonance wavelengths based on the perturbation-resonance interactions; see the representative peak-shift-based modality in Fig. 1C]. Instead, non-resonance amplification is a universal phenomenon of signal amplification that exists at arbitrary wavelength, but it is noticeable only after the generation of an EC (we will describe this in detail in the following sections). The non-resonance amplification results in significantly less Joule heating and enables the operation at various wavelengths for a given device. Using our proposed method with 785-nm wavelength light on a low cost microscope that nominally has extremely poor contrast and signal-to-noise ratio (SNR), we successfully visualized the shapes and positions of perturbations with features as small as a 25-nm radius in a 63-nm wide region. This demonstrates the unprecedented effectiveness, robustness, and simplicity of the framework. We believe, our work paves the route to revolutionize the US$ 1.8-billion market of conventional diffraction-limited optical microscopes.

**Physical models of ECs and non-resonance amplification**

To explain the visualizable non-resonance nanoscale detection, let's consider a pair of parallel nanowires with identical material and topology as shown in Fig. 2A. A scarlet colored cuboid representing a nanoscale perturbation approaches the nanowires and thus perturbs the electromagnetic modes. Here, we do not assume any specific type of material or topology for the perturbation; thus, in practice, it can represent many types of objects including viruses, nanoparticles, supermolecules, environmental dust particles, or even fabrication errors on the nanowires. An eigenmode expansion analysis combined with coupled mode perturbation theory *(19-22)* shows that the electric field $\mathbf{E}_{sp}(\mathbf{r})$ around the symmetry plane (SP) (see the transparent plane marked with "SP" in Fig. 2, A and B) can be represented by (see Sec. 1 in the supplement)

$$\mathbf{E}_{SP}(\mathbf{r}) = \mathbf{E}_{SP}(\mathbf{r}, c_1^{k_i}) + \mathbf{E}_{SP}(\mathbf{r}, c_2^{k_i}) + \mathbf{E}_{SP}(\mathbf{r}, \Delta c_1^{k_i}\big|_{\text{pert}}) + \mathbf{E}_{SP}(\mathbf{r}, \Delta c_2^{k_i}\big|_{\text{pert}}) + \mathbf{E}_{SP}(\mathbf{r}, c_3^{u}), \quad (1)$$

where $c_1^{k_i}$ and $c_2^{k_i}$ are unperturbed eigenmode expansion coefficients corresponding to the first and second nanowires, respectively. $\Delta c_1^{k_i}\big|_{\text{pert}}$ and $\Delta c_2^{k_i}\big|_{\text{pert}}$ are the perturbation coefficients representing the impact of the perturbation on the first and second nanowires, respectively. $c_3^{u}$ is the coupling matrix with respect to the perturbation. Hence, $\mathbf{E}_{SP}(\mathbf{r}, c_1^{k_i}) + \mathbf{E}_{SP}(\mathbf{r}, c_2^{k_i})$, $\mathbf{E}_{SP}(\mathbf{r}, \Delta c_1^{k_i}\big|_{\text{pert}}) + \mathbf{E}_{SP}(\mathbf{r}, \Delta c_2^{k_i}\big|_{\text{pert}})$, and $\mathbf{E}_{SP}(\mathbf{r}, c_3^{u})$ can be physically interpreted as the field contributions induced by the nanowires, the coupling between the perturbation and the nanowires, and the perturbation itself, respectively. For conventional excitation modes (a single incoherent or coherent excitation beam) in a brightfield or darkfield microscope, the transverse area of the nanowires is much smaller than the beam size. Therefore, the local gradient of the electric field can be neglected and the pair of nanowires are excited analogously and re-radiate in an identical manner on the SP in the nearfield region, i.e., $\mathbf{E}_{SP}(\mathbf{r}, c_1^{k_i}) = \mathbf{E}_{SP}(\mathbf{r}, c_2^{k_i})$. A permanent constructive interference is thus formed on the SP. Because the dimensions of nanowires are much larger than those of the perturbation, the second and third terms: $\mathbf{E}_{SP}(\mathbf{r}, \Delta c_1^{k_i}|_{pert}) + \mathbf{E}_{SP}(\mathbf{r}, \Delta c_2^{k_i}|_{pert}) + \mathbf{E}_{SP}(\mathbf{r}, c_3^u)$ relating to the perturbation are overwhelmed by the constructive interference $\mathbf{E}_{SP}(\mathbf{r}, c_1^{k_i}) + \mathbf{E}_{SP}(\mathbf{r}, c_2^{k_i})$ from the background. This is one of the reasons why conventional low-SNR optical microscopy is unable to detect nanoscale perturbations. However, if we can somehow excite the pair of nanowires into an anti-symmetric state, *i.e.*, $\mathbf{E}_{SP}(\mathbf{r}, c_1^{k_i}) = -\mathbf{E}_{SP}(\mathbf{r}, c_2^{k_i})$, we would create an EC because the dominant field $\mathbf{E}_{SP}(\mathbf{r}, c_1^{k_i}) + \mathbf{E}_{SP}(\mathbf{r}, c_2^{k_i})$ would disappear. In the EC, only the field contribution from the perturbation-related terms: $\mathbf{E}_{SP}(\mathbf{r}, \Delta c_1^{k_i}|_{pert}) + \mathbf{E}_{SP}(\mathbf{r}, \Delta c_2^{k_i}|_{pert}) + \mathbf{E}_{SP}(\mathbf{r}, c_3^u)$ would remain. Equally important, the coupling terms $\mathbf{E}_{SP}(\mathbf{r}, \Delta c_1^{k_i}|_{pert})$ and $\mathbf{E}_{SP}(\mathbf{r}, \Delta c_2^{k_i}|_{pert})$ do not cancel out provided that the perturbation is not exactly in the middle of the pair of nanowires. Therefore, we can engineer the shape and dimensions of the nanowires such that the total perturbation signal is significantly amplified, i.e.,

$$\left| \mathbf{E}_{SP}(\mathbf{r}, \Delta c_1^{k_i}|_{pert}) + \mathbf{E}_{SP}(\mathbf{r}, \Delta c_2^{k_i}|_{pert}) + \mathbf{E}_{SP}(\mathbf{r}, c_3^u) \right|^2 \gg \left| \mathbf{E}_{SP}(\mathbf{r}, c_3^u) \right|^2. \qquad (2)$$

This aforementioned phenomenon can be understood by analogy as the constructive and destructive interference between two wave trains that move in opposite directions; see the schematics shown in Figs. 2A and B. The detectability of a perturbation relies on the fact that the strength of scattering is larger than the measurement noise and error, $\varepsilon$, of the detection system.

For a low-performance detection system or an extremely small perturbation, the uncoupled scattering from the perturbation alone $|\mathbf{E}_{SP}(\mathbf{r}, c_3^u)|^2$ is usually dominated by $\varepsilon$. However, we can use non-resonance amplification to make the signal detectable, i.e.,

$$\left| \mathbf{E}_{SP}(\mathbf{r}, \Delta c_1^{k_i}|_{pert}) + \mathbf{E}_{SP}(\mathbf{r}, \Delta c_2^{k_i}|_{pert}) + \mathbf{E}_{SP}(\mathbf{r}, c_3^u) \right|^2 \gg \varepsilon. \qquad (3)$$

The reason we call $\left| \mathbf{E}_{SP}(\mathbf{r}, \Delta c_1^{k_i}|_{pert}) + \mathbf{E}_{SP}(\mathbf{r}, \Delta c_2^{k_i}|_{pert}) + \mathbf{E}_{SP}(\mathbf{r}, c_3^u) \right|^2$ non-resonance amplification is that it is a universal phenomenon that happens at arbitrary wavelengths (see the derivations in the supplement, in which we did not impose any assumptions on the wavelength), whereas resonator sensors only work around their resonant wavelengths. We should emphasize that the non-resonance amplification also exists in the case with symmetric excitation [ $\mathbf{E}_{SP}(\mathbf{r}, c_1^{k_i}) = \mathbf{E}_{SP}(\mathbf{r}, c_2^{k_i})$ ], but as discussed, the background scattering from the nanowires $\left| \mathbf{E}_{SP}(\mathbf{r}, c_1^{k_i}) + \mathbf{E}_{SP}(\mathbf{r}, c_2^{k_i}) \right|$ is much stronger than that from the perturbation. Hence, the generation of an EC is the prerequisite for the application of non-resonance amplification in the perturbation detection. The EC generated by the anti-symmetric excitation has an extraordinary feature, i.e., it exists regardless of the gap size between the pair of nanowires. This is easy to understand because in the mathematical derivations, we did not impose any assumption for the gap, which means the destructive interference $\mathbf{E}_{SP}(\mathbf{r}, c_1^{k_i}) = -\mathbf{E}_{SP}(\mathbf{r}, c_2^{k_i})$ always holds under anti-symmetric excitation provided that the nanowires have identical material and topology. In fact, the generation of EC can also be elegantly explained with dipolar approximation for the pair of nanowires. A systematic study on the EC based on two-dipole interference can be found in Sec. 2 of the supplement, in which we have numerically validated this feature of ECs. A paradigm-shifting result is that we can fabricate nanowire-based non-resonance sensors with a small

footprint (size is limited only by the fabrication method) along the direction perpendicular to the SP.

It is clear now that the key is to generate an EC (via anti-symmetric excitation), such that the non-resonance amplification is detectable. The insets on the bottom right corners of Fig. 2A and B are the brightfield images obtained by nearfield computation and Fourier optics for symmetric (conventional) and anti-symmetric excitations *(23, 24)*, respectively. See Sec. 3 in the supplement for more details about the simulation methods. Apparently, one can clearly observe an EC between the pair of nanowires and find the perturbation (marked by "P") from the elongated left pattern in the inset of Fig. 2B, whereas one can only find the information of the nanowires in the symmetric excitation case shown in the inset of Fig. 2A.

**Implementations: configuration and modelling**

One may ask, how can we excite the pair of nanowires anti-symmetrically especially for a sub-diffraction-limited gap using macroscale illumination? In fact, two-beam interference, which is widely used in optical interferometry provides the answer; however, the operation mode will use totally different physics in our case. As schematically shown in Fig. 2C, two in-phase $y$-polarized plane waves from distant locations along the $\pm$ $x$-axis impinge at oblique angles on a sample located near the origin. They interfere to produce the standing wave electric field excitation $\mathbf{E}_s = E_0 \cos(2\pi x/\Lambda)\hat{\mathbf{y}}$ with periodic phase jumps, where $\Lambda = \lambda/\sin\theta$ is the interference period. Note the period for intensity is $\Lambda/2$. Thus, we can excite the anti-symmetric state by positioning the pair of nanowires at $x = (m + 1/2) \Lambda/2 \pm p$, respectively, where $m$ is any integer and $2p$ is the center-to-center spacing of the nanowires. See the bottom two insets relating to the intensity and phase distribution of the standing waves in Fig. 2C, which shows the pair of nanowires being anti-symmetrically excited at a representative position.

Based on the two-beam anti-symmetric excitation, we first use simulation to investigate the field enhancement $\left|\mathbf{E}_{SP}(\mathbf{r},\Delta c_1^{k_i}\big|_{pert})+\mathbf{E}_{SP}(\mathbf{r},\Delta c_2^{k_i}\big|_{pert})+\mathbf{E}_{SP}(\mathbf{r},c_3^u)\right|^2$ of various perturbations around a pair of nanowires with fixed dimensions and compare the results with those of conventional brightfield and darkfield imaging. Figure 3A shows schematics of the three imaging modalities. For fixed dimensions shown in the inset on the top left corner of Fig. 3B, changing the materials of nanowires for the sensing of a $SiO_2$ perturbation does not alter the conclusion that the perturbation-related field using the proposed framework is orders of magnitude stronger than the fields from conventional brightfield and darkfield imaging modalities. The same trend remains if we fix the material of nanowires to be Si but vary both the material ($SiO_2$, Si, Ag, and $TiO_2$ perturbations) and dimension of the perturbation; see Fig. 3C. We then move the nanoscale perturbation to different positions around the nanowires and compute the far-field images, by which we can clearly observe the positions of perturbation in a sub-diffraction-limited volume (see Fig. 3D and 3E). Note that the gap between nanowires is far smaller than the diffraction limit of the imaging system. The two insets in the bottom right corners of the top subfigures in Figs. 3D and 3E are optical images without perturbations, which are used for comparison. An exception is that if the perturbation is located exactly in the middle of the pair of nanowires (see the third subfigure in Fig. 3E), one cannot find the perturbation from the image because the coupling terms $\mathbf{E}_{SP}(\mathbf{r},\Delta c_1^{k_i}\big|_{pert})$ and $\mathbf{E}_{SP}(\mathbf{r},\Delta c_2^{k_i}\big|_{pert})$ cancel out. This exactly shows the symmetric nature of EC. Because the SP is only a single line in the sample space (see the dotted line in the third subfigure of Fig. 3E), and because the shapes of real perturbations are usually irregular, the probability of failing to see a perturbation is negligible.

We should remind our readers not to confuse the proposed framework with structured illumination microscopy (SIM) *(25)*, because our proposal relies on the localized phase jump in the standing wave, while SIM does not. A direct evidence is that SIM will result in a resolution improvement not exceeding a factor of two, while the anti-symmetric excitation will result in the generation of an EC regardless of the gap size (see Figs. S1-S4 in the supplement). We should mention that a pair of nanowires is not the exclusive choice, but any other pair of identical nanostructures with arbitrary geometries can be utilized to generate the EC (see Fig. S7 in the supplement). This opens up a space for engineering the non-resonance sensor such that $\left| \mathbf{E}_{SP}(\mathbf{r}, \Delta c_1^{k_i}\big|_{pert}) + \mathbf{E}_{SP}(\mathbf{r}, \Delta c_2^{k_i}\big|_{pert}) + \mathbf{E}_{SP}(\mathbf{r}, c_3^u) \right|^2$ can be maximized at a given excitation wavelength. The reason why we choose nanowires is that they can excite strong bright modes when the polarization of the illumination beam is parallel to the long axis of nanowires *(26, 27)*.

**Experiments with diverse perturbations**

We fabricated multiple double-nanowire and quad-nanowire structures (the nominal width is 50 nm) with diverse perturbations using electron beam lithography (EBL); see Fig. S10 in the supplement. Because the nanowire structures are widely used in the semiconductor industry, we can use them to mimic the detection of perturbations in typical intentional defect wafers *(3)*. The nanowire structures can also be embedded in liquid environment to sense free-form perturbations like viruses and supermolecules, because liquids do not influence the formation of the EC other than to change the interference period $\Lambda$ due to the increased refractive index. We then constructed a two-beam far-field interference system using a single-mode single-frequency 785-nm wavelength laser with integrated optical isolator to implement the proposed idea. We use a top-down microscope with a low numerical aperture objective (0.4 NA) and a low-performance 14-megapixel camera to capture 726-μm × 582-μm field of view images of the nanowire structures with perturbations. The schematic and picture of the detection system are shown in Fig. 4A and Fig. S8, respectively. Because the system is operated in the widefield mode, the field-of-view is only limited to the aperture of the system and size of camera. See a representative full field of view image of the sample captured by the top-down microscope in Fig. 4B. We crop and zoom these images to clearly show selected regions of interest in Figs. 4C-E and Fig. 5. Only after completing the optical imaging do we collect scanning electron microscope (SEM) images to corroborate our findings in these regions. The classical Abbe limit of the microscope is 981 nm ($\lambda$/2NA). However, because of the extremely low performance of the imaging system (see more details in the supplement), only a 4-µm gap can be roughly resolved (see Figs. S9B and C in the supplement) from the brightfield image and the scattering of an isolated nanoscale object (390 nm × 120 nm, in Fig. S9D) can hardly be found from the darkfield image. This demonstrates again that we cannot use conventional brightfield and darkfield imaging modalities to sense nanoscale perturbations, let alone in a deep sub-wavelength cavity.

The first step to demonstrating our proposal is to validate the generation of ECs using two-beam interference. Because the upper and lower nanowires need to be positioned at $x = (m + 1/2) \Lambda/2 \pm p$, respectively, we fix the two-beam illumination apparatus and scan the sample along the direction perpendicular to the long axis of nanowires such that the EC with a best contrast can be found. Figure 4C shows the SEM image of the fabricated double-nanowire structure that has an 80-nm gap. The sample's orientation is controlled by a rotating stage (RS) to ensure the long axis of the nanowires is parallel to the polarization orientation (y-direction) of the field. Figure 4C shows the best-focal image of the double-nanowire structure as the sample is translated along $x$ (i.e., the gap is moved relative to the positions of the intensity nulls). The purple ribbon delineates the central region of bright spots from the outside sidelobes. We say that an EC has been generated if there is an intensity minimum at the center of the purple ribbon. Figure 4C shows that the double-nanowire structure undergoes EC and non-EC transitions in a repeated manner as it is moved. The single-nanowire structure, however, does not form an EC regardless of position; Fig. 4D shows that its image always has one bright spot. Figure 4E shows that we can generate ECs for double-nanowire structures with various gaps. As a comparison, we present the darkfield and brightfield images of the same double-nanowire structures captured by the same imaging system; see Figs. 4F and 4G. Apparently, we cannot visualize the gaps and shapes of the double-nanowire assembly from either of the conventional image sets. The slight fabrication imperfections in some of the nanowires (see the SEM images in Fig. 4E) can be directly visualized in the optical images in Fig. 4E. This directly demonstrates the sensitivity of our system to perturbations.

We now consider the direct imaging of a diverse set of intentional nanoscale perturbations. Because the NA of the objective is only 0.4, we choose the 14-μm long nanowires such that the corresponding optical images are "lines" (not "dots," like that for 2-μm long nanowires). This facilitates the observation and classification of the perturbations. The first sample is a quad-nanowire structure with a tiny dot (represented by "a" in Fig. 5A) positioned in between the upper nanowires, to mimic a typical semiconductor defect or a nanoparticle. See the 3D schematic with the dot marked in red in the left panel of Fig. 5A. An SEM measurement shows that the diameter of the dot is around 50 nm. From Fig. 5A, we can clearly and directly observe the perturbing dot in the optical image when the EC is created. When no EC is created, the dot is buried by the background constructive interference. The second sample is a double-nanowire structure with a large perturbation ($\delta_a$ = 90 nm, where $\delta$ is the width of the perturbation; see the inset in the 3D schematic of Fig. 5B showing the definition of $\delta$) and a small perturbation ($\delta_a$ = 50 nm) on the bottom left and bottom right corners, respectively. Moreover, the bottom nanowire is much narrower (32 nm) than the top one. From the Fig. 5B optical image with the EC, we can clearly observe a brighter spot in the bottom left corner corresponding to the larger perturbation, and the bottom pattern apparently has a much weaker intensity than that of the top pattern. These details are not visible in the non-EC image. Figure 5C presents a quad-nanowire structure with two central line-expansion perturbations "a" and "b." The dimension of this type of perturbation can be characterized by the expanded width $\Delta$; see the inset in the 3D schematic of Fig. 5C showing the definition of $\Delta$. Similar to those in Figs. 5A and 5B, one can find two bright spots with the brighter one corresponding to the larger perturbation ($\Delta_b$ = 31 nm) in only the EC optical image. A more complicated quad-nanowire structure, shown in Fig. 5D, has tilted 63-nm gaps and two central bumps that break the field symmetry. One can clearly resolve the corners with gaps (marked with "a", "c", "e", and "f" in Fig. 5D) from the EC optical image. Moreover, because the left central bump is thicker than the right one [marked by "b" and "e"], the excited bright modes are stronger on the left half of the anti-symmetric optical image compared to on its right half. These images clearly demonstrate that diverse perturbations, including the isolated, bonded, expanded, and defective ones, can all be detected. We should emphasize that all the edge-to-edge gaps (750 nm, 188 nm, 137 nm, and 63 nm) between the upper and lower nanowires are smaller than the diffraction limit 981 nm. This validates our derivation that an EC can be generated regardless of the gap size. Thus, the sensing of perturbations can be implemented in a nanometric region. We validated the proposal using semiconductor perturbations that are widely seen in integrated devices *(3)*. One can seamlessly apply the same prototype for the visualizable sensing of nanoparticle and biomaterial (e.g., viruses or molecule clusters) by patterning functional groups around nanowires. Once target analytes are trapped, they act as perturbations that are directly visualized from the optical images. See the Fig. S11 artwork.

**Conclusion**

We have experimentally visualized nanoscale perturbations by introducing a simple add-on illumination apparatus to create an electromagnetic canyon combined with a chip-scale field amplifier consisting of nanowires that amplify the perturbation signals. Compared to existing detection modalities, the proposed framework has five major advantages. First, it enables the direct imaging of nanoscale perturbation in a sub-diffraction-limited volume using a conventional microscope without any post-processing whatsoever. Second, the nanowire-based amplifier is easy to fabricate using a single EBL and dry-etching step. Third, the method uses non-resonance amplifiers to boost the SNR of the perturbation; thus, the prototype can be operated at various wavelengths and without generating as much Joule heating. Fourth, the total device width can be as small as 250 nm (limited only by our EBL) while the length can be selected to best suit the intended application. Fifth, the method uses a widefield microscope and thereby enables arbitrarily large fields of view (limited only by the camera sensor size and the ability to fabricate and align arrays of nanowires to the illumination pattern) to be observed in a single-shot frame. Although we validate our framework with a low-performance microscope and nanowires with relatively large variations, an instrument with higher performance (for instance, a camera with a deeper electron well depth) and a nanofabrication technique with better resolution is recommended for the cases where detecting even smaller nanoscale perturbations is required. In conclusion, we believe our work opens up a new avenue for visualizable nanoscale detection and sensing. The technique enables optical microscopy to solve challenging problems across many fields including semiconductor defect inspection, microelectronics testing, biosensing, material characterization, virus counting, and microfluidic monitoring.

**Acknowledgments:** The authors gratefully acknowledge Cisco System Inc. for access to its Arcetri cluster. L.L.G. acknowledges the Center for Advanced Study at the University of Illinois for teaching release time. This work was funded by Cisco Systems Inc. (Gift Awards CG 1141107 and CG 1377144), University of Illinois at Urbana-Champaign College of Engineering Strategic Research Initiative, and Zhejiang University – University of Illinois at Urbana-Champaign (ZJUI) Institute Research Program.

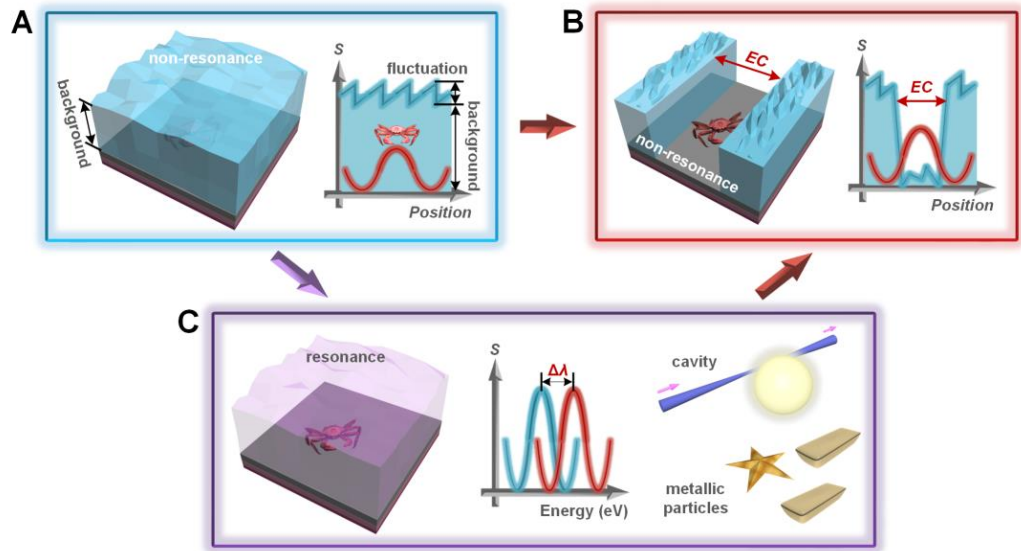

**Fig. 1. Artworks and schematics for the concept of visualizable nanoscale detection with an electromagnetic canyon.** (**A**) Artwork showing the difficulty in sensing nanoscale objects (represented by a crab) in conventional imaging modalities because of the sea of electromagnetic background signals (represented by blue water) and fluctuations (represented by sawtooth waves). (**B**) Artwork showing the proposed framework of visualizable nanoscale sensing by artificially creating an electromagnetic canyon. (**C**) Artwork and schematic showing the mechanisms of resonance-based (including mode spectrum of microcavities and localized surface plasmon resonance in metallic particle systems) nanoscale sensing.

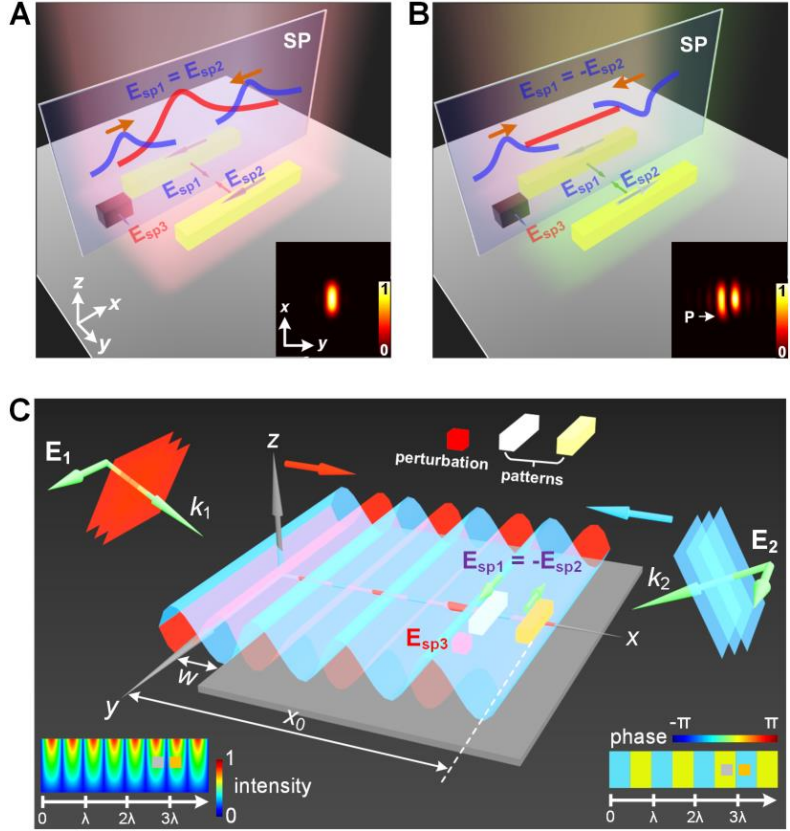

**Fig. 2. Simplified schematics for showing the mathematical principle and experimental implementation of the proposed framework of nanoscale sensing**. (**A** and **B**) Nanostructure ensemble consisting of a pair of dielectric nanowires and a perturbation on a substrate under (**A**) conventional and (**B**) anti-symmetric excitation modes. Each nanowire is 20 nm wide by 800 nm long by 60 nm tall. The edge-to-edge gap for the nanowires is 100 nm. The perturbation is a short nanowire that is 20 nm wide by 100 nm long by 60 nm tall. The distance between the center of the perturbation and symmetry plane (SP) is 30 nm. The insets at the bottom right-hand corner are the corresponding bright-field images. $\mathbf{E}_{sp1}$ and $\mathbf{E}_{sp2}$ denote the unperturbed electric field on the SP arising from the left and right nanowires. $\mathbf{E}_{sp3}$ is the simplified form of the last term on the right-hand side of Eq. (1). The self-coupling among the elements in the ensemble is not shown. (**C**) Generation of the anti-symmetric excitation modes at many different locations for the nanostructure ensemble along the sample using two-wave interference with inclined angles. The insets at the bottom left and right corners of **C** are the intensity and phase of the standing wave versus *x*. The different colours (white and yellow) used for the pair of nanowires in **C** indicate that the system is excited into the anti-symmetric mode.

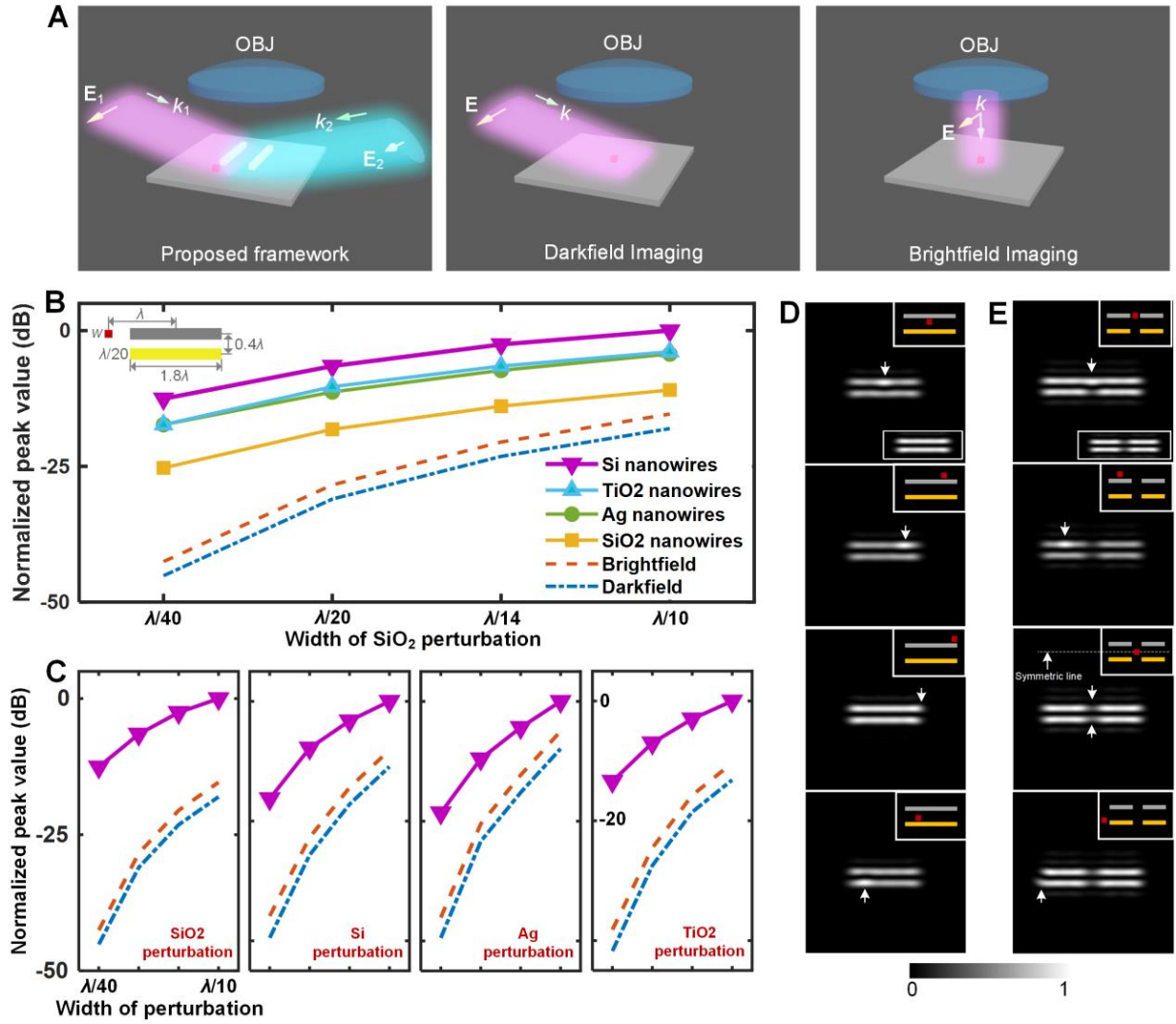

**Fig. 3. Simulated non-resonance amplification for perturbations with diverse materials and dimensions**. (**A**) Schematics showing the setups of the proposed framework, darkfield imaging, and brightfield imaging. (**B**) Simulated far-field signals versus the size of a SiO$_2$ perturbation. Different nanowire materials are shown for the proposed approach. (**C**) Investigation of the impact of the dimension and the material of the perturbation on the detection signal for the three imaging modalities. The dimensions and material (Si) of the nanowires are fixed for the proposed approach. See the schematic for the dimensions of the nanowires in the top left corner of Fig. 3B. (**D** and **E**) Simulated far-field images for a nanoscale perturbation at different positions around a double-nanowire and a quad-nanowire structure, respectively. The center-to-center spacing of the nanowires is 0.4$\lambda$, which is smaller than the theoretical diffraction limit 0.61$\lambda$ of the simulated imaging system. For the simulation of brightfield images, we only consider the scattering from nanowires and the perturbation, while the background reflection from the substrate is removed. The different colors of nanowires are for representing that they are out-of-phase.

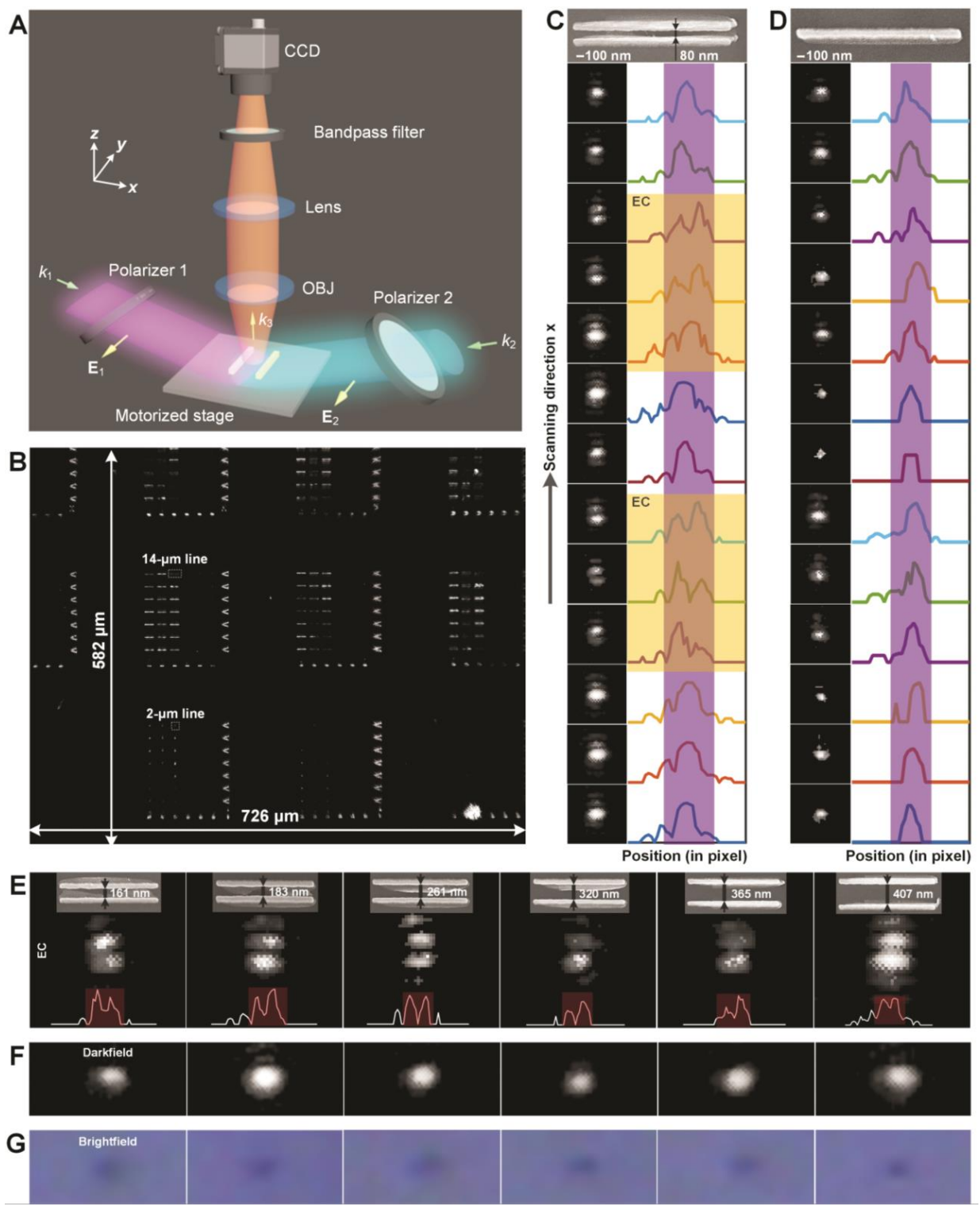

**Fig. 4. Experimental setup and results.** (**A**) Schematic of the two-beam interference system. (**B**) A representative wide-field image captured by the system of the investigated sample that consists of various patterns. The two tiny dotted boxes underneath the word "μm" in the labels

show typical regions of interest for subsequent parts of Fig. 3. (**C** and **D**) Zoomed-in images of the double-nanowire and single-nanowire structures captured at various scan positions (100-nm increment) along the x-axis. The curves on the right part of (**C**) and (**D**) are the corresponding slices along the x-axis. The portions of the curves that are surrounded by a purple ribbon are the main lobes. The curves surrounded by a yellow ribbon correspond to the positions where an EC is formed because anti-symmetric excitation applies. (**E**), Zoomed-in images of the ECs formed using two-beam interference on double-nanowire structures with various gaps. (**F-G**), Corresponding darkfield and brightfield zoomed-in images of the double-nanowire structures.

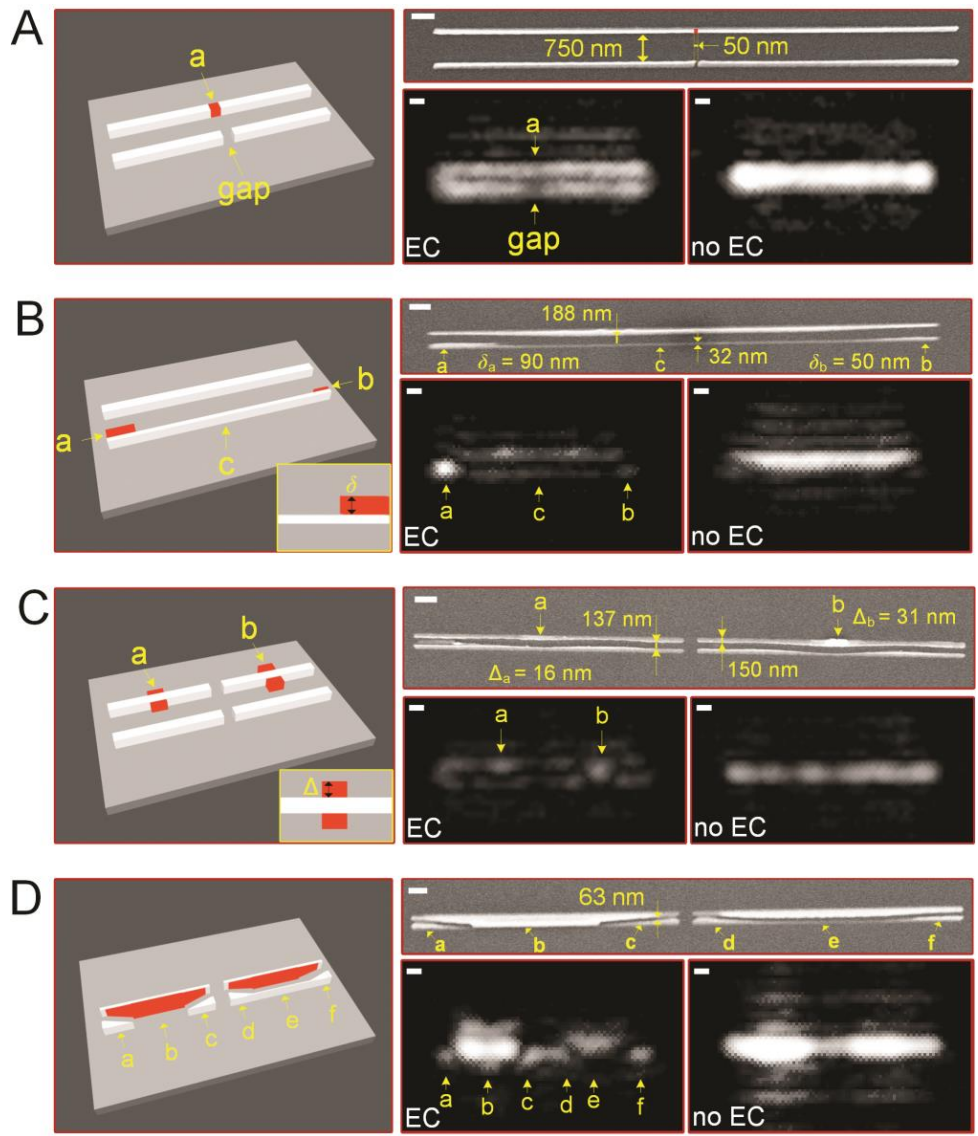

**Fig. 5. Experimental imaging for diverse perturbations**. (**A-D**) Experimental images of patterns with various perturbations for the anti-symmetrically excited and symmetrically excited cases. All of the gaps in **A-D** are smaller than the nominal 981-nm resolution limit (actual resolution is 4 µm because of noise) of the top-down microscope. Perturbations are marked in red in **A-D**. The scale bars in both the SEM images and the optical images in **A-D** are 500 nm. The nominal width of nanowires is 50 nm.

# Supplementary Materials for

## Visualizable Detection for Nanoscale Perturbations

Jinlong Zhu[1,*,†], Aditi Udupa[1], Lynford L. Goddard[1,*]

Correspondence to: zhuwdwz1@illinois.edu

# Materials and Methods

## 1. Mathematical model for generating an electromagnetic canyon

The pair of nanowires and the perturbation form an electrostatic ensemble whose modes can be found by first solving a self-sustained eigenvalue problem for each individual nanostructure *(28)*

$$\sigma_i(\mathbf{r}) = \frac{\gamma_i}{2\pi} \oint \sigma(\mathbf{r}_i) \frac{(\mathbf{r}-\mathbf{r}_i)\cdot\vec{\mathbf{n}}_i}{|\mathbf{r}-\mathbf{r}_i|^3} dS_i \text{ and}$$
$$\tau_i(\mathbf{r}) = \frac{\gamma_i}{2\pi} \oint \tau(\mathbf{r}_i) \frac{(\mathbf{r}_i-\mathbf{r})\cdot\vec{\mathbf{n}}_i}{|\mathbf{r}-\mathbf{r}_i|^3} dS_i \quad (i=1,2,3),$$

(S1)

where $\sigma_i$, $\tau_i$, and $\vec{\mathbf{n}}_i$ are the charge, dipole, and normal vector at position $\mathbf{r}$ on the surface of the $i^{th}$ nanostructure. $\gamma_i = (\varepsilon_i - \varepsilon_{\text{eff}\_i}) / (\varepsilon_i + \varepsilon_{\text{eff}\_i})$ is the eigenvalue associated with the corresponding eigenvalue equation. $\varepsilon_i$ is the electric permittivity of the $i^{th}$ nanostructure, and $\varepsilon_{\text{eff}\_i}$ is the effective background permittivity obtained by considering the substrate-induced "image nanostructure" *(17)*. Once the materials and dimensions are given, the surface charge eigenmodes $\sigma_i^k(\mathbf{r})$ for $k = \{1, 2, ..., \infty\}$ can be uniquely determined and the surface charge distribution can be represented by a superposition of these eigenmodes. Note that the eigenfunctions describe the intrinsic properties of a self-sustained system. Thus, they are valid irrespective of the external illumination source used and are only determined by the materials and topology. The electric field at an arbitrarily spatial position $\mathbf{r}$ from the ensemble is given by

$$\mathbf{E}(\mathbf{r}) = \sum_{i=1}^{M} \oint \mathbf{g}(\mathbf{r}-\mathbf{r}_i) \sum_{k_i=1}^{\infty} c_i^{k_i} \sigma_i^{k_i} dS_i \quad (i=1,2,3),$$

(S2)

where $\mathbf{g}(\mathbf{r}-\mathbf{r}_i)$ is the vectorial Green's function, and $c_i^{k_i}$ is the $k_i^{th}$ coefficient to be determined for the $k_i^{th}$ eigenmode $\sigma_i^{k_i}$ for the $i^{th}$ nanostructure. $M$ denotes the number of isolated nanostructures in the ensemble. Here $M$

= 3. The excitation of these modes in a nanostructure is determined by both the external sources and the evanescent fields due to the nearby nanostructures. It is therefore the undetermined coefficients in the eigenmode expansion for the surface charge and for the surface dipole distribution that include the effects from the self-coupling among the nanostructures of the ensemble. The expansion coefficient $c_i^{k_i}$ can be described via the biorthogonality between $\sigma_i^{k_i}$ and $\tau_i^{k_i}$ (20)

$$c_i^{k_i} = \beta_i^{k_i} \oint \tau_i^{k_i} \vec{\mathbf{n}}_i \cdot \left[ \mathbf{E}_{i0} + \sum_{p=1}^{M} \mathbf{E}_{ip} \right] dS_i, \quad (S3)$$

where $\beta_i^{k_i} = \dfrac{(\varepsilon_i - \varepsilon_{\text{eff}\_i})(\varepsilon_i^{k_i} - \varepsilon_{\text{eff}\_i})}{\varepsilon_i^{k_i} - \varepsilon_i}$ and $\varepsilon_i^{k_i}$ is the real permittivity associated with the $k_i^{\text{th}}$ eigenvalue of the $i^{\text{th}}$ nanostructure. $\mathbf{E}_{i0}$ and $\mathbf{E}_{ip}$ respectively denote the applied field and the electric field arising from the $p^{\text{th}}$ nanostructure at position $\mathbf{r}_i$ on the surface of the $i^{\text{th}}$ nanostructure. Here, the electric field from the $i^{\text{th}}$ nanostructure acting on itself is assumed to be zero. If the observation point $\mathbf{r}$ in Eq. (S2) is on the surface of the $j^{\text{th}}$ nanostructure, the expansion coefficient can be obtained by combining Eqs. (S2) and (S3) followed by algebraic operations

$$c_i^{k_i} = \xi_{i0}^{k_i} + \sum_{h=1}^{\infty} \sum_{l=1}^{M} \sum_{m=1}^{M} \sum_{d=1}^{\infty} f_{ilm}^{k_i hd} \xi_{m0}^d, \quad (S4)$$

where $f_{ilm}^{k_i hd} = \left( \delta^{k_i h} \delta_{il} - C_{li}^{hk_i} \right)^{-1} C_{lm}^{hd}$ and $\delta$ denotes the Kronecker delta function. $\xi_{i0}^{k_i}$ and $\xi_{m0}^d$ are given by

$$\begin{aligned} \xi_{i0}^{k_i} &= \beta_i^{k_i} \oint \tau_i^{k_i}(\mathbf{r}_i) \vec{\mathbf{n}}_i \cdot \mathbf{E}_{i0} dS_i \text{ and} \\ \xi_{m0}^d &= \beta_m^d \oint \tau_m^d(\mathbf{r}_m) \vec{\mathbf{n}}_m \cdot \mathbf{E}_{m0} dS_m, \end{aligned} \quad (S5)$$

where $\mathbf{E}_{i0}$ and $\mathbf{E}_{m0}$ denote the applied field on the surfaces of the $i^{\text{th}}$ and $m^{\text{th}}$ nanostructures, respectively. Here, we should mention that the subscripts $i$ and $m$ do not have any essential difference; the only difference is that $i$ denotes the quantity to be determined, while $m$ denotes the quantity to be superposed in Eq. (S4) arising from the self-coupling within the ensemble. $C_{li}^{hk_i}$ is the term that describes the coupling from the $i^{\text{th}}$ particle to the $l^{\text{th}}$ particle

$$C_{li}^{hk_i} = \beta_l^h \oint \oint \tau_l^h(\mathbf{r}_l)\vec{\mathbf{n}}_l \cdot \mathbf{g}(\mathbf{r}_l - \mathbf{r}_i)\sigma_i^{k_i}(\mathbf{r}_i)\mathrm{d}S_l \mathrm{d}S_i. \tag{S6}$$

$C_{lm}^{hd}$ can be derived analogously by simply updating the subscripts and superscripts. If the bright modes can be excited via external sources *(21, 22, 29)*, the nanostructure radiates out propagation waves into the surrounding background and thus the effect of radiation damping should be considered in the electrostatic interaction. For deep sub-wavelength objects, dipolar radiation dominates the oscillation damping; thus, higher-order multipoles can be neglected. Accordingly, the expansion coefficient in Eq. (S4) can be modified by adding the damping terms *(20, 29)*

$$c_i^{k_i} = \xi_{i0}^{k_i} + j\alpha_i P_i^{k_i} + \sum_{h=1}^{\infty}\sum_{l=1}^{M}\sum_{m=1}^{M}\sum_{d=1}^{\infty} f_{ilm}^{k_i hd}\left(\xi_{m0}^d + j\alpha_m P_m^d\right), \tag{S7}$$

where $\alpha_i = k^3/6\pi\varepsilon_{\mathrm{eff}\_i}$ and $\alpha_m = k^3/6\pi\varepsilon_{\mathrm{eff}\_m}$. The dipole term $P_i^{k_i}$ is given by

$$P_i^{k_i} = \beta_i^{k_i} \oint \tau_i^{k_i}(\mathbf{r}_i)\vec{\mathbf{n}}_i \cdot \mathbf{p}_i \mathrm{d}S_i, \tag{S8}$$

where $\mathbf{p}_i$ is the dipole moment of the $i^{\mathrm{th}}$ nanostructure. $P_m^d$ is given by Eq. (S8) with an updated superscript. Knowing the expansion coefficients, the electric field at an arbitrary position $\mathbf{r}$ can be computed via Eq. (S2) provided the eigenfunctions are predetermined.

As shown in Fig. 2, the left and right nanowires are identical in both their material and their topology; thus, the corresponding eigenfunctions are normally identical, *i.e.*, $\sigma_1^{k_i} = \sigma_2^{k_i}$ and $\tau_1^{k_i} = \tau_2^{k_i}$. Accordingly, the elements of the coupling matrix described in Eq. (S6), would normally satisfy $C_{12}^{hk_i} = C_{21}^{hk_i}$ and $C_{11}^{hk_i} = C_{22}^{hk_i}$. Here, we should recall the fact that the coupling matrices do not depend on the external sources. In conventional excitation where the transverse area of the nanostructures is much smaller than the beam size, the local gradient of the electric field can be neglected. Thus, the field that excites the left and right nanowires can be regarded as identical to each other, *i.e.*, $\mathbf{E}_{10} = \mathbf{E}_{20}$; see Fig. 2(a). This indicates that $\xi_{10}^d = \xi_{20}^d$. Due to the geometrical symmetry without the nanoscale perturbation, $P_1^{k_i} = P_2^{k_i}$ because of $\mathbf{p}_1 = \mathbf{p}_2$. Hence, we have $c_1^{k_i} = c_2^{k_i}$ and the electric field at an arbitrary symmetric plane (SP) from the pair of nanowires is always the result of constructive interference. When the perturbation is present, the symmetry is broken, and the elements of coupling matrix are changed by the coupling between the perturbation and the pair of nanowires, *i.e.* $c_i^{k_i}\big|_{\text{pert}} = c_i^{k_i} + \Delta c_i^{k_i}\big|_{\text{pert}}$ $(i = 1, 2)$, where $\Delta c_i^{k_i}\big|_{\text{pert}}$ denotes the change and is given by

$$\Delta c_i^{k_i}\Big|_{\text{pert}} = j\alpha_i \left( P_i^{k_i}\Big|_{\text{pert}} - P_i^{k_i} \right) + \sum_{h=1}^{\infty}\sum_{l=1}^{3}\sum_{m=1}^{3}\sum_{d=1}^{\infty} f_{ilm}^{k_i hd} \left( \xi_{m0}^d + j\alpha_m P_m^d\Big|_{\text{pert}} \right)$$
$$- \sum_{h=1}^{\infty}\sum_{l=1}^{2}\sum_{m=1}^{2}\sum_{d=1}^{\infty} f_{ilm}^{k_i hd} \left( \xi_{m0}^d + j\alpha_m P_m^d \right) \quad (i = 1, 2), \tag{S9}$$

where $P_m^d\big|_{\text{pert}}$ denotes the perturbed dipole distribution. The electric field on the SP then can be represented as

$$\mathbf{E}_{\text{SP}}(\mathbf{r}) = \mathbf{E}_{\text{SP}}(\mathbf{r}, c_1^{k_i}) + \mathbf{E}_{\text{SP}}(\mathbf{r}, c_2^{k_i}) + \mathbf{E}_{\text{SP}}(\mathbf{r}, \Delta c_1^{k_i}\big|_{\text{pert}}) + \mathbf{E}_{\text{SP}}(\mathbf{r}, \Delta c_2^{k_i}\big|_{\text{pert}}) + \mathbf{E}_{\text{SP}}(\mathbf{r}, c_3^u), \tag{S10}$$

where $c_3^u$ is the coupling matrix with respect to the perturbation. Each term in the right-hand side of Eq. (S10) has an explicit expression that can be easily obtained by substituting $c_i^{k_i}\big|_{\text{pert}} = c_i^{k_i} + \Delta c_i^{k_i}\big|_{\text{pert}}$ $(i = 1, 2)$ followed by an expansion according to the serial number of particles. From the previous discussion, we recall that $\mathbf{E}_{\text{SP}}(\mathbf{r}, c_1^{k_i})$ and $\mathbf{E}_{\text{SP}}(\mathbf{r}, c_2^{k_i})$ do not cancel out but instead form a strong constructive interference on the SP. If the dimension of the perturbation is much smaller than that of the nanowires, the terms [ $\mathbf{E}_{\text{SP}}(\mathbf{r}, \Delta c_1^{k_i}\big|_{\text{pert}}) + \mathbf{E}_{\text{SP}}(\mathbf{r}, \Delta c_2^{k_i}\big|_{\text{pert}}) + \mathbf{E}_{\text{SP}}(\mathbf{r}, c_3^u)$ ] related to the perturbation can be overwhelmed by the constructive interference. Hence, it is difficult to directly observe the nanoscale perturbations from the far-field images, which is especially the case when the pair of nanowires has a sub-wavelength gap. If we excite the nanowires into an anti-symmetric state, *i.e.*, $\mathbf{E}_{20} = -\mathbf{E}_{10}$, we have $\xi_{10}^{k_i} = -\xi_{20}^{k_i}$, $\xi_{10}^d = -\xi_{20}^d$, $P_1^{k_i} = -P_2^{k_i}$, and $P_1^d = -P_2^d$. This indicates $c_1^{k_i} = -c_2^{k_i}$ and thus the unperturbed terms $\mathbf{E}_{\text{SP}}(\mathbf{r}, c_1^{k_i})$ and $\mathbf{E}_{\text{SP}}(\mathbf{r}, c_2^{k_i})$ cancel out each other on the SP in the nearfield region, leaving only the perturbation-related terms. This means we may observe the nanoscale perturbation even if the investigated perturbation is much smaller than the wavelength.

## 2. The role of symmetry in generating an electromagnetic canyon: dipolar approximation

Consider a pair of closely positioned identical dielectric nanoscale objects on the *x*-axis at *xyz* coordinates (-*p*, 0, 0) and (*p*, 0, 0), where $p \ll \lambda$. Assume that they are illuminated by a monochromatic beam propagating along the *z*-direction in an epi-illumination microscope. Typically, in classical imaging, the excitation is described as "beam-like," i.e., the field has a constant or a slowly-varying cross-section, e.g., a plane wave or a Gaussian beam, respectively. When the objects are located in the path of a collimated beam (wide-field imaging) or at the beam's focus (confocal imaging), they are simultaneously and isotropically excited and form the symmetric polarization state. This is equivalent to two in-phase electric dipoles of equal amplitude; see Fig. S1A. In contrast, consider anisotropically exciting the objects and forming the anti-symmetric polarization state. The two dipoles now have equal amplitudes but oscillate perfectly out-of-phase. Figure S1B vividly depicts this using two arrows with opposite orientations. Using the electric dipole approximation and starting from equation 9-18 of Ref. 23, we can express the complex amplitude $\mathbf{E}_t(\mathbf{r})$ at an arbitrary observation point $\mathbf{r}$ in both cases as:

$$\mathbf{E}_t(\mathbf{r}) = \xi_1 \mathbf{D}_1 + \xi_2 \mathbf{D}_2 + \gamma_1 (\mathbf{D}_1 \cdot \hat{\mathbf{r}}_1) \hat{\mathbf{r}}_1 + \gamma_2 (\mathbf{D}_2 \cdot \hat{\mathbf{r}}_2) \hat{\mathbf{r}}_2, \tag{S11}$$

where $\mathbf{D}_1$ and $\mathbf{D}_2$ are the complex amplitudes of the dipoles, $\mathbf{D}_1(t) = \text{Re}[\mathbf{D}_1 e^{-i\omega t}]$ and $\mathbf{D}_2(t) = \text{Re}[\mathbf{D}_2 e^{-i\omega t}]$. $\hat{\mathbf{r}}_1$ and $\hat{\mathbf{r}}_2$ are the unit vectors originating from the left and right dipoles, respectively, and going to the point **r**. $\xi_j$ and $\gamma_j$ ($j$ = 1, 2) are complex coefficients that depend on the distance between the observation point and the source, i.e.,

$$\xi_j = \frac{k^2 e^{ikr_j}}{4\pi\varepsilon_0 r_j}\left[1 - \frac{1}{k^2 r_j^2} + \frac{i}{kr_j}\right]$$
$$\gamma_j = \frac{k^2 e^{ikr_j}}{4\pi\varepsilon_0 r_j}\left[\frac{3}{k^2 r_j^2} - \frac{3i}{kr_j} - 1\right], \ j = 1, 2. \tag{S12}$$

Here, $k$ is the wavenumber, $\varepsilon_0$ is the permittivity of vacuum, and $r_j$ denotes the distance from the $j^{\text{th}}$ dipole to the arbitrary observation point. The first two and last two terms on the right-hand side of Eq. (S11) of the main text can be interpreted as the dipole-induced and the position offset-induced field contributions, respectively, i.e., $\mathbf{E}_d = \xi_1\mathbf{D}_1 + \xi_2\mathbf{D}_2$ and $\mathbf{E}_r = \gamma_1(\mathbf{D}_1\cdot\hat{\mathbf{r}}_1)\hat{\mathbf{r}}_1 + \gamma_2(\mathbf{D}_2\cdot\hat{\mathbf{r}}_2)\hat{\mathbf{r}}_2$. We denote the phase difference between $\mathbf{D}_1$ and $\mathbf{D}_2$ as $\alpha$, where $0 \leq \alpha \leq \pi$. Thus, $\alpha$ equals 0 and $\pi$ for the cases in Figs. S1A and S1B, respectively.

The symmetry properties of the $x$, $y$, and $z$ electric field components in Eq. (S11) across the $x = 0$ and the $y = 0$ planes are the key factors in determining whether the electromagnetic canyon (EC) can be generated. Moreover, $\alpha$ determines these symmetry properties. The symmetric state results in constructive interference and a merging of the objects in the image while the anti-symmetric state creates a destructive interference "splitting-line" (i.e., the EC) across which we can perfectly resolve the two objects. Table 1 summarizes the symmetry properties of each field component for different dipole excitations. To start, let us compare symmetric and anti-symmetric $y$-polarized dipoles. For these cases, $E_y$ is the dominant near-field component; see Fig. S2. For symmetric $y$-polarized dipoles, $E_y$ is symmetric across both planes and thus it is not possible to resolve the dipoles in the microscope image; see Fig. S1C. The same conclusions hold for symmetric $x$-polarized and $z$-polarized dipoles (not shown). However, for anti-symmetric $y$-polarized dipoles, $E_y$ is anti-symmetric in $x$. Thus, there is a near-field splitting line for $E_y$ and the objects are resolved in the microscope image; see Fig. S1J. Moreover, $E_x$ is anti-symmetric in $y$ while $E_z$ is anti-symmetric in both $x$ and $y$. Anti-symmetry across at least one plane for each near-field component produces a perfect destructive interference in the microscope image at the intersection of the two planes. Thus, the microscope image will have zero intensity at the center point $(x, y) = (0, 0)$. See Fig. S3A, which shows the two dipoles are resolved with infinite contrast ratio. Contrast ratio is defined as the ratio of the peak intensity in the image to the value at $(0, 0)$. A paradigm-shifting consequence of the anti-symmetry in the excited dipole moments is an EC with perfect null. Diffraction and interference in the lens-based system now play the role of catalyst instead of barrier for creating high contrast ratio images. For anti-symmetric $x$-polarized and $z$-polarized dipoles, we also obtain a splitting line and anti-symmetry across $x = 0$ for the near-field maps of $E_x$ and $E_z$, respectively. See Fig. S3. However, the symmetry across both planes of $E_z$ for $x$-polarized dipoles and of $E_x$ for $z$-polarized dipoles causes the contrast ratio to decrease to approximately 10 (a slightly imperfect but usable EC) and 1 (no EC), respectively; see Fig. S3B and S3C. Whereas $z$-polarized dipoles do not form an EC, the $x$-polarized dipoles still form the canyon because the near-field longitudinal component $E_z$ has a weaker effect on the far-field microscope image than the near-field transverse components, $E_x$ and $E_y$. To generate the EC, $\alpha$ need not be exactly 180°, i.e., tolerance exists. We can excite the objects into a partially anti-symmetric state. See Figs. S1D-I, where we can clearly see an EC for two objects spaced by $d = 2p = \lambda/4$ when $\alpha \geq 150°$. We herein define a gap-dependent threshold $\alpha_t(d)$ to be the smallest value of $\alpha$ for which the two objects form an EC, i.e., a local minima exists in the middle of the two peaks. Figure S1 shows that $\alpha_t(\lambda/4) = 150°$.

Under anti-symmetric excitation, not only can we generate the EC, but we can observe this EC using a microscope objective with an arbitrarily small NA; compare Fig. S4B with Fig. S4A. This is because the microscope objective's NA does not change the symmetry properties of the fields. For both the *x*-polarized and *y*-polarized anti-symmetric cases, when the gap, 2*p*, is reduced below Abbe's limit, the peak intensity begins to drop off rapidly (approximately as $p^2$), but we should mention that the dropped intensity belongs to the pair of objects (nanowires in Fig. 2), not to a perturbation (if present). However, the apparent object gap (AOG), i.e., the gap in the image space divided by the system magnification, converges to a constant after the expected decrease in the regime of geometrical optics. See Figs. S4C and S4D. The overlap of the curves for *x*- and *y*-polarization indicates that an arbitrary linear combination of the transversally anti-symmetric polarization does not significantly alter the formation of the EC for the two objects; see Figs. 4E and 4F. If $\mathbf{D}_1$ and $\mathbf{D}_2$ include anti-symmetric *z*-components, the microscope image degenerates, but there is still a local minimum (i.e., an EC is formed) if the magnitude of the longitudinal component of the dipole moments does not exceed three times that of the transverse component; see Fig. S5. The above results and conclusions directly extend to polychromatic (i.e., white-light) excitation that is anti-symmetric for the dominant wavelengths, which builds a solid foundation for experimental validations. Note that the amplitude and absolute phase of each wavelength can be arbitrary.

## 3. Details of the imaging setup in simulations

The coherent optical imaging microscope used in the simulation for Figs. 2 and 3 is in a widefield configuration with a 100× magnification and a 0.8 input numerical aperture (NA) of the objective. The output NA of the imaging optics is chosen as 1. To compute the vectorial images of nanostructures, three procedures, *i.e.*, the definition of sources, the computation of near-field, and the propagation of EM field for imaging, are implemented.

**3.1 Definition of sources and near-field computation**

The near-field of the nanostructure assembly used to compute the images are obtained via the finite-different time-domain method. For a pair of nanowires positioned on top of a substrate, in order to anti-symmetrically excite them, we use the standing waves formed by two-beam interference with an inclined angle of 60º, as proposed in Fig. 2C. However, we should remind our readers that two-beam interference is not the only way to achieve anti-symmetric excitation; any vectorial beam that has a feature of a local $\pi$-shift in phase can be utilized for anti-symmetric excitation, provided the pair of objects is positioned symmetrically about the null. A near-field observation plane that is above the investigated area is used to record the scattering field from the excited nanostructure assemble. The area of the observation plane should be large enough to allow most of the time-averaged backward scattering power (>95%) to flow through.

**3.2 Near-field decomposition and imaging**

The electric field captured by the observation plane is decomposed into a series of plane waves using far-field projection, after which the plane wave components within the input NA are focused onto the image plane by chirped z-transform *(24)*. This imaging methodology is naturally succinct and fast. To evaluate its accuracy, we applied another imaging method that is referred to as the equivalent magnetic-dipole (EMD) based vectorial electromagnetic field imaging *(30-32)*. In this method, the electric field on the observation plane is decomposed into many EMDs followed by the ray tracing for all the field lines of EMDs within the input NA using the generalized Jones matrix formalism. The Debye-Wolf integral then can be applied to calculate the image of all the EMDs. If the observation plane is large enough [to allow most of the time-averaged backward scattering power (>95%) to flow through] and if the sampling interval for EMDs is small enough ($\lambda/3$ in our estimation), the EMD based method can give an accurate image for an arbitrary target. The drawback of the EMD based method is that it is time-consuming if the number of EMDs is very large. For the computation of the image with 150 × 150 observation points corresponding to 301 × 301 EMDs, it takes 13.3 minutes in MATLAB integrated programming environment on an in-house built workstation with two Intel Xeon E5-2683 v3 2.0 GHz 28-core processors. However, by parallelizing the program onto 16 Pacini computing nodes of Cisco's Arcetri cluster using C++, the computation time reduced to only 8.4 s. As the EMD based method naturally inherits the vectorial properties of EM imaging, it is a reasonable benchmark to estimate the chirped z-transform based method. Accordingly, we compute the microscope images on the best focal plane for a dipole pair (with $\lambda/15$ gap) that is polarized along *y*-direction with anti-symmetry about $x = 0$; see Fig. S6. The intensity curves corresponding to the central cross-sections of the two images from both methods highly overlap with only a minor mismatch on the sidelobes. Hence, the time-efficient chirped z-transform based imaging method is accurate enough to give reasonable images for the investigated samples under the current wide-field configuration.

## 4. Generation of electromagnetic canyons with various objects

We now consider the generation of ECs with various real objects, where the finite dimension-induced non-uniform polarization and influence of the background cannot be neglected. The mathematical model developed in Sec. 1 of the supplemental indicates that any pair of objects can be utilized to generate the EC. The first object is a silicon bowtie on top of a $SiO_2$ substrate. The gap size (20 nm) is smaller than many common bowtie structures *(33, 34)*. The left and right triangles are impinged by a *y*-polarized transverse beam with phases -π/2 and π/2, respectively.

Each beam for anti-symmetric excitation is a localized plane wave (see the inset at the top left corner in Fig. S7E). This is the simplest and most generalized method to excite the anti-symmetric state. The bright-field microscope captures images of the bowtie area in epi-illumination mode. Figure S7A shows that the peaks for the left and right triangles are clearly resolved and that an EC has been generated in between. The $SiO_2$ substrate, which essentially plays the role of "mirror bowtie," does not alter the anisotropic polarization but only changes the strength of electric polarization. In transmission mode, the top background behaves as the "mirror material." Figure S7B shows the result for nanoholes etched on an infinite thin plate. Researchers often use the double-nanowire structure as the "gold standard" for evaluating a microscope's resolving power. Figures S7C and S7D show clear ECs for epi-illumination and transmission mode, respectively. To further understand tolerances, we explored the effect of reducing the excitation amplitude on the left side and the effect of adding line edge roughness (LER) to the right-side of the double-nanowire structure. Figures S7E and S7F show that both the excitation-induced and the material-induced breaking of the perfectly anti-symmetric polarization, indeed disturb the images. The result is an asymmetric pattern in the image space, but still the ECs are clearly generated even when the amplitude of the left excitation is only 40% of that of the right excitation or when a $\lambda/13$-scale LER exists.

## 5. Experimental setup

The light source is a fiber-coupled continuous-wave single-mode single-frequency 785-nm wavelength laser (Thorlabs FPV785S) that is split into two output ports using a 50:50 fiber splitter (Thorlabs TN785R5F2). A 1-inch diameter 25-mm focal length lens collimates each output into a free-space beam. Each beam passes through a rotatable free-space linear polarizer (Thorlabs WP25M-UB). Two fiber polarization controllers (Thorlabs FPC020) are inserted in the fiber path between the fiber splitter output ports and the collimating lenses in order to maximize and equalize the powers of the two beams that are transmitted through the free-space polarizers. Two high-precision rotation mounts (Thorlabs PRM1) are utilized to control the inclined angles of illumination beams with a resolution of 5 arcmin. The beams then interfere to form an anti-symmetric excitation field on the sample. Using a long coherence length laser, matching the optical path lengths, and carefully aligning the angular orientations of the two beams were critical elements for obtaining the desired anti-symmetric field and thereby generating the ECs. One beam can be blocked to create a conventional illumination field by tuning the fiber polarization controller. See the picture of the two-beam anti-symmetric excitation apparatus in Fig. S8A.

The sample is mounted on a manual rotation stage (Thorlabs RP01) on top of an *xyz* translation stage (Thorlabs PT3) with motorized actuators (Thorlabs ZST225B) and is scanned relative to the interference field, typically with steps of size 100 nm. The microscope is a simple 4-f imaging system consisting of a Mitutoyo M Plan Apo NIR 20X objective with 0.4 NA, an ordinary 1-inch diameter 75-mm focal length lens (resulting in 7.5X magnification), and an Amscope MU1403B 4096 × 3286 pixel CMOS camera with its built-in infrared filter removed to allow the 785-nm light to reach the camera sensor. A 780-nm bandpass filter (Thorlabs FBH780-10) was inserted to reduce the effect of stray room light. Before the sensing experiments, a white-light LED (Thorlabs LEDWE-50) is utilized (by the control of a home-made switch box) to image the sample and locate the patterned areas on the wafer. A full-view of the entire experimental system with the marked optical components in the imaging path is shown in Fig. S8C. The entire setup was covered with a plastic cover (not shown) to minimize noise from air drafts in the room.

The two-beam interference apparatus shown in Fig. S8A is aligned by viewing the overlapping shape of the two beams projected on a viewing card until a minimized area of the spot is achieved for a given inclined angle. See the series of pictures showing the alignment process and the beam spot observed in a darkened room in Fig. S9A. The entire imaging path is assembled without any image correction algorithms or hardware. In an imaging system, besides the imperfections in optical components, the noise sources (especially the shot noise) in the camera contribute a significant portion to the distortion, low contrast, and small SNR of captured images. We use a low-cost CMOS camera (MU1403B; see Fig. S8B) with the dynamic range and maximal SNR as low as 65.3 dB and 35.5 dB, respectively, to validate the robustness of the proposed framework. The observation of structures with deep subwavelength features using conventional imaging modalities is not possible. The imaging performance of the built microscope system is tested with two patterns on a NIST 8820 artifact *(35, 36)*. Apparently, the images suffer from extremely low contrast and small SNR, and only the 4-µm gap can be barely resolved, although the theoretical resolution limit is 981 nm. See Figs. S9B and C. Moreover, we also cannot detect the information of an isolated nanostructure (390 nm × 120 nm) from the darkfield imaging mode; see the darkfield image and SEM measurement of the nanostructure show in Fig. S9D.

## 6. Sample fabrication

Rogue Valley Microdevices fabricated four-inch diameter, 525-µm thick, p-type, Si wafers with a 6-µm thick wet thermal oxide and 150-nm thick stoichiometric nitride layer deposited by low pressure chemical vapor deposition (LPCVD). This layer structure is not important but is presented here for completeness. We cleaved one such wafer into 2.2 cm × 2.2 cm square pieces. Next, we performed electron beam lithography using PMMA resist, deposited 5 nm of Ti and 100 nm of Au, and performed metal liftoff to define single-nanowire, double-nanowire, and quad-nanowire structures with nominal design gaps ranging from 0 nm to 1000 nm with 50-nm increments. We then deposited a blanket coating of 2 nm of Au to make the top of the wafer electrically conductive. This ensures that we can take SEM images of the exact same sample immediately after optical imaging. See the "Graphic Data System" (GDS) file together with a full SEM view of the entire patterns in Figs. S10A and S10B. The actual gaps, denoted in the SEM images, are slightly different than the designed dimensions due to the proximity effect and other fabrication imperfections. See the SEM images of two representative double-nanowire structures in Figs. S10C and S10D with nanoscale variations.

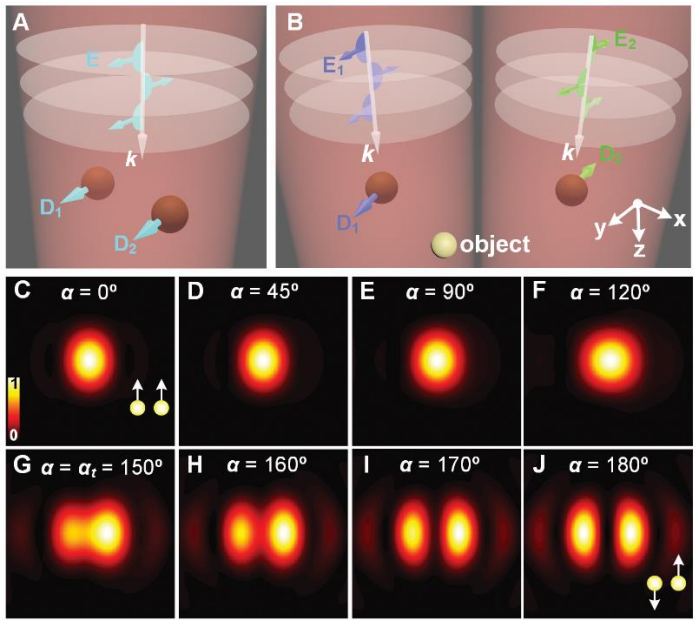

Figure S1: Comparison of symmetric and anti-symmetric excitation for creating an EC using a pair of point objects. (**A** and **B**) Schematic showing symmetric and anti-symmetric illumination. The arrows on each object denote the excited dipolar moments. (**C-J**) Microscope intensity images of the pair of dipoles with a gap of $\lambda/4$ as the function of $\alpha$. The insets on the bottom right corner of **C** and **J** denote the in-phase (symmetric) and out-of-phase (anti-symmetric) excitation states. The field of view is 1 μm × 1 μm in the sample space.

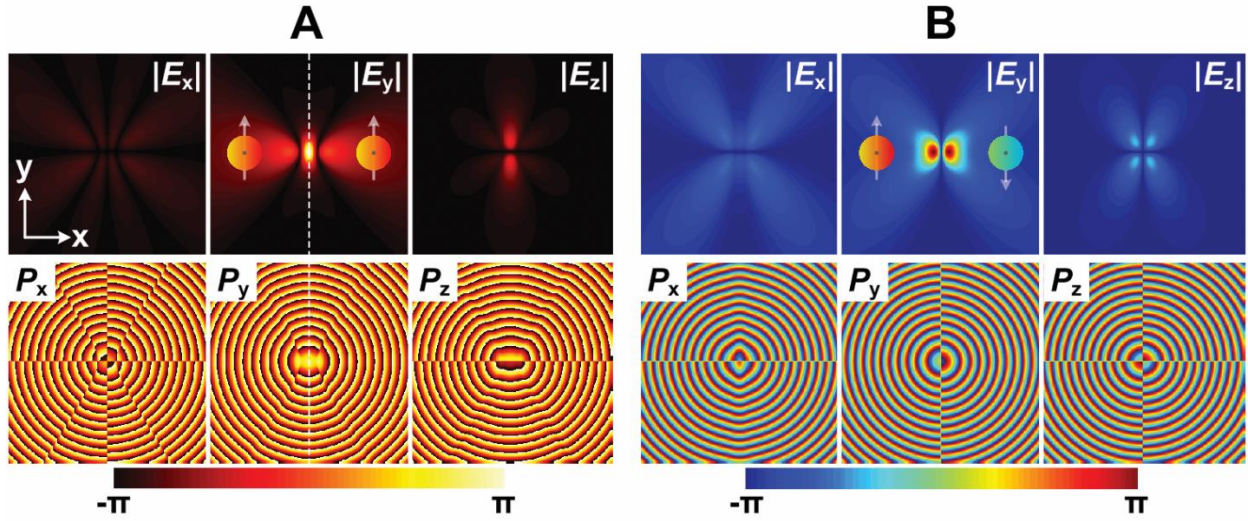

Figure S2: Comparison of the near-fields for symmetric and anti-symmetric *y*-polarization. (**A** and **B**) Amplitude and phase distributions of different field components on a near-field plane that is 4-μm wide by 4-μm long and 1 μm above the *x-y* plane. The gap for the pair of dipoles is $\lambda/4$.

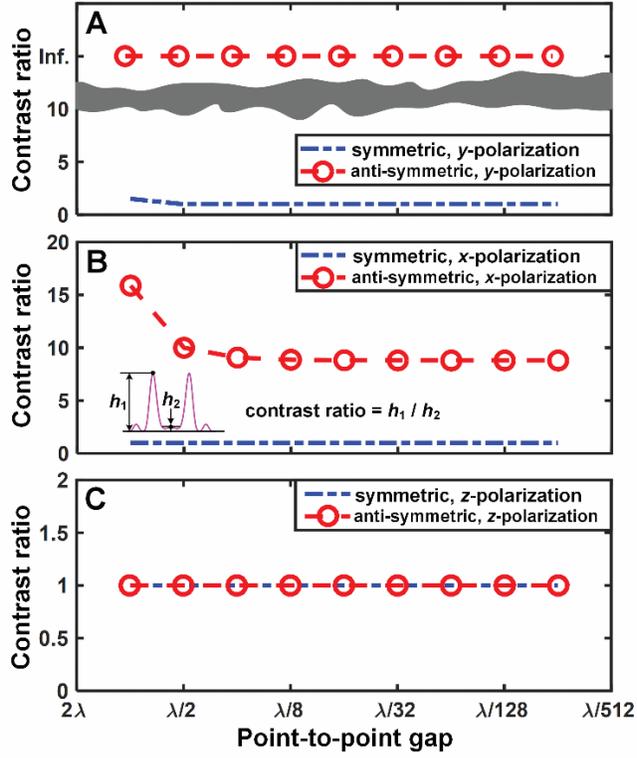

Figure S3: Contrast ratio as a function of the point-to-point gap. (**A**) *y*-polarization. (**B**) *x*-polarization. (**C**) *z*-polarization. The red dashed and blue dotted curves denote the contrast ratio for the anti-symmetric and symmetric dipoles, respectively. The inset on the bottom left corner of **B** shows the definition of the contrast ratio, which is the ratio of the intensity at the brightest spot in the image to the intensity at the center of the image and is a measure of the quality of the EC.

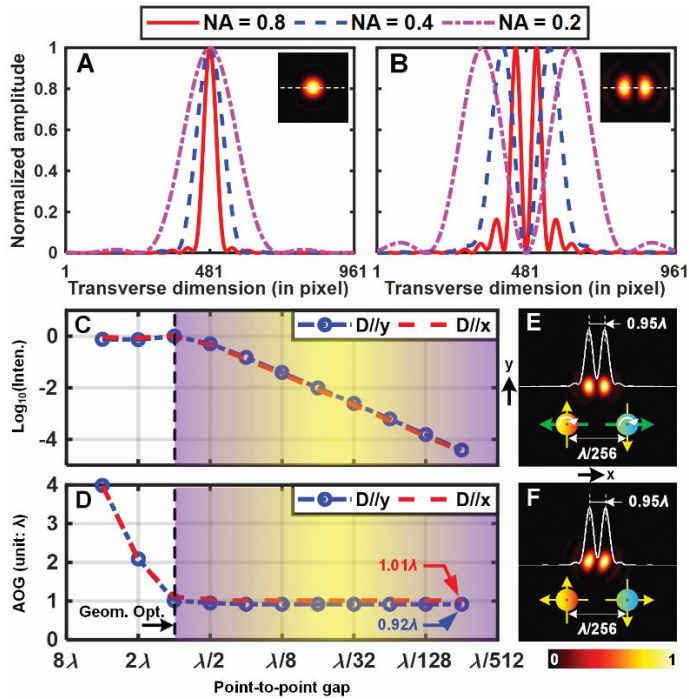

Figure S4: Roles of anisotropic *x*-polarization and *z*-polarization in generating an EC from a pair of point objects. (**A** and **C**) Field vectors on the SP for the anti-symmetric *x*-polarized and *z*-polarized dipoles, respectively. (**B** and **D**) Bright-field microscope images (top) and the dominant components of the electric field and their phase distribution maps in the near-field region for the anti-symmetric *x*-polarized and *z*-polarized dipoles, respectively. The amplitude and phase distribution maps are computed on an observation plane that is 4 μm wide by 4 μm long and 1 μm above the *x*-*y* plane. The gap between the pair of dipoles is 0.25$\lambda$.

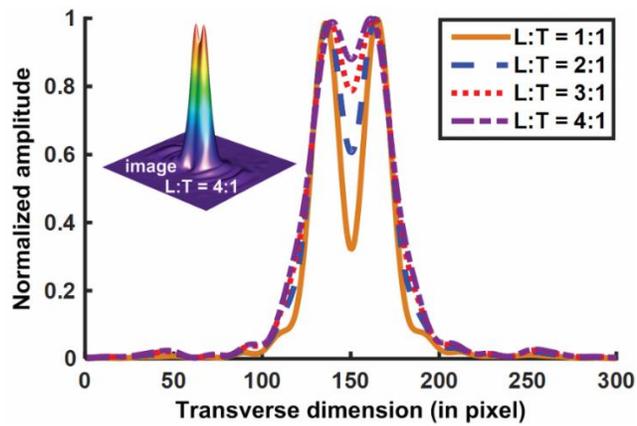

Figure S5: Effect of the longitudinal component of a pair of electric dipoles on the generation of the EC. The gap for the pair of dipoles is $\lambda/4$. L: amplitude of the longitudinal component; T: amplitude of the transverse component. The magnification and input NA of the wide-field microscope are 100× and 0.8, respectively. Each pixel on the microscope image is 2 μm × 2 μm, which corresponds to 20 nm × 20 nm in the sample space.

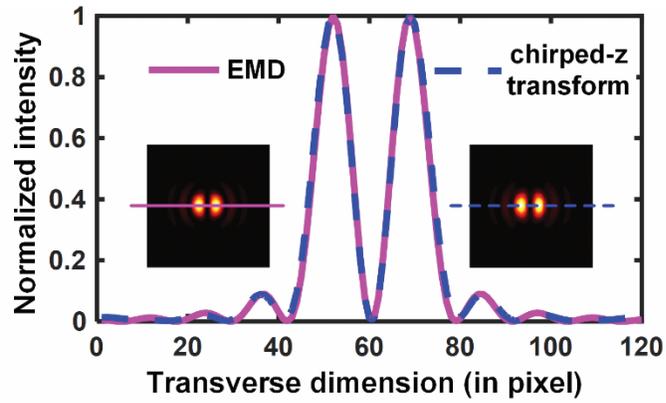

Figure S6: Validation for the imaging methods. Computed microscope images (normalized) and the central slices from the EMD and chirped-$z$ transform based imaging methods. The field of view for both insets is 4 μm × 4 μm in the sample space. Each pixel on the microscope image is 2 μm × 2 μm, which corresponds to 20 nm × 20 nm in the sample space. The gap for the pair of dipoles is $\lambda/15$.

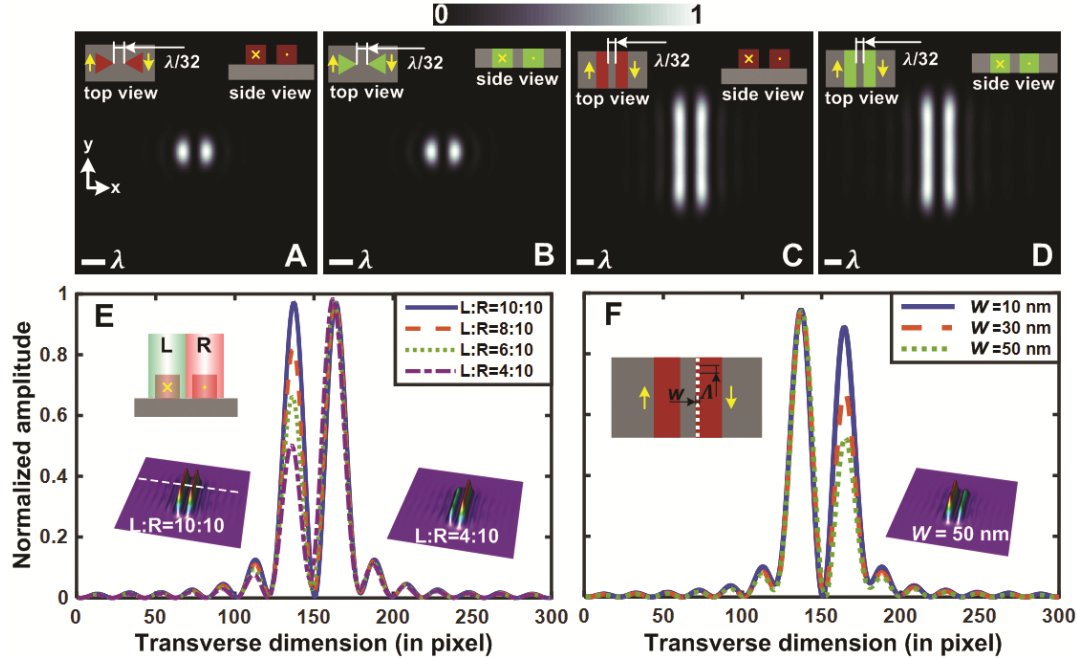

Figure S7: Creation of ECs using anti-symmetric excitation of bowtie and double-nanowire structures. (**A** to **B**) Bright-field images of the bowtie in reflection and transmission modes, respectively. (**C** to **D**) Corresponding images for the double-nanowire structure. The edge-to-edge distance is $\lambda/32$ for all cases, and the side length of the bowtie and the width of the lines are $\lambda/10$ and $\lambda/13$, respectively. The field of view of (**A** to **D**) is 4 μm × 4 μm in the sample space. (**E** and **F**) Effect of nonuniform illumination and LER on the quality of the generated EC for the double-nanowire. The central cross-sections for the bright-field images under various right-side amplitudes of **E** excitation strength and **F** LER. L: amplitude of the left-side excitation; R: amplitude of the right-side excitation. $w$: width of the LER; $\Lambda$: pitch of the LER. The gap and duty cycle are fixed at 80 nm and 0.5, respectively. The magnification and input numerical aperture of the bright-field microscope are 100× and 0.8, respectively. Each pixel on the microscope image is 2 μm × 2 μm, which corresponds to 20 nm × 20 nm in the sample space.

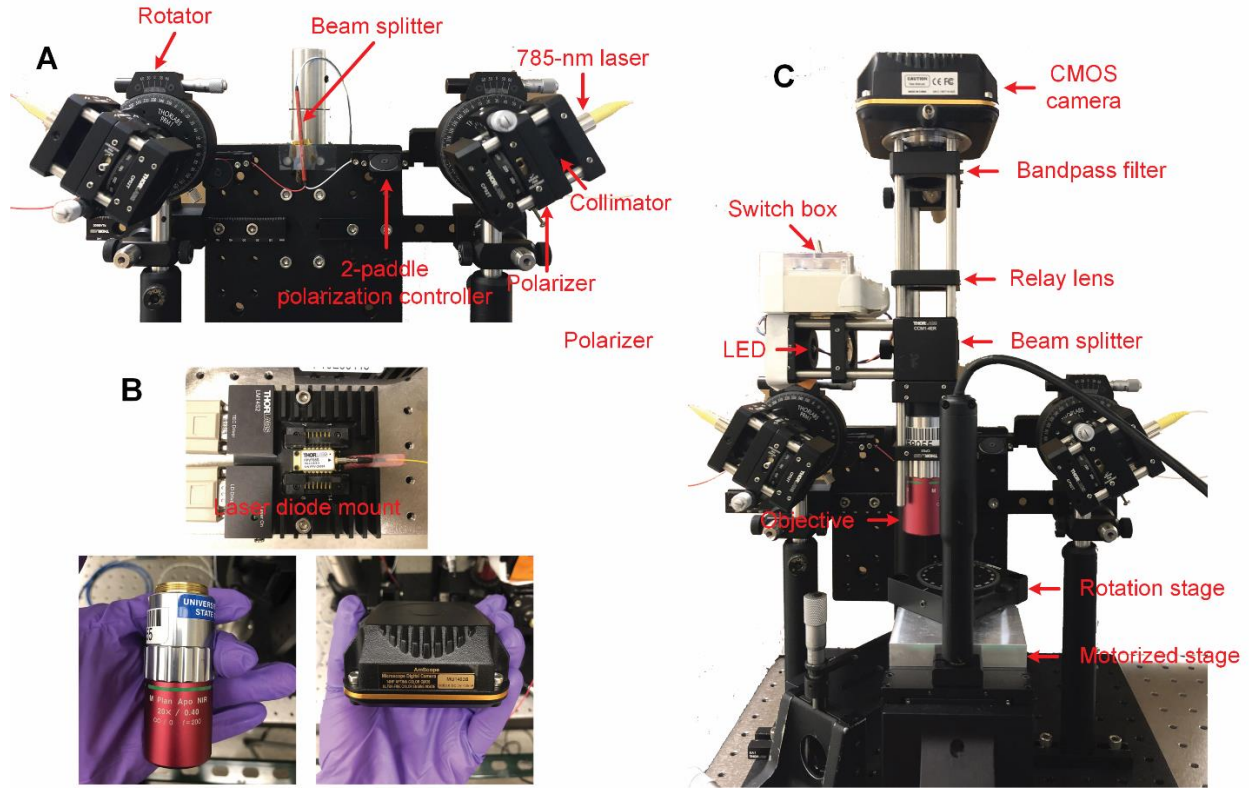

Figure S8: Experimental systems with primary components marked out. (**A**) The two-beam interference apparatus for generating the ECs. (**B**) The 785-nm laser diode mount, the NIR objective, and the low-cost CMOS camera used in the system. (**C**) Full view picture of the entire experimental system including the two-beam interference apparatus and the top-down widefield microscope.

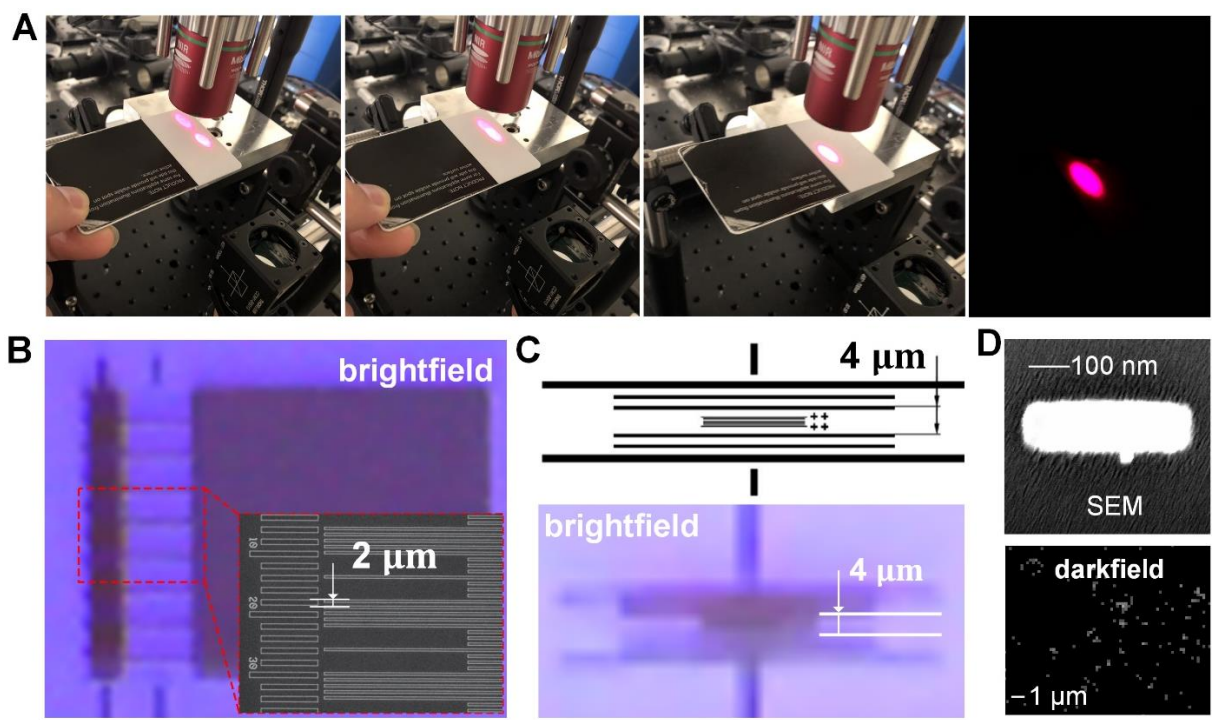

Figure S9: Calibration and testing for the experimental system. (**A**) Images during alignment for showing the merging process of the two beams. The merged beam spot with a minimal area taken in a darkened room is shown on the very right panel of **A**. (**B**) Optical image captured by the top-down widefield microscope under conventional white-light illumination for a region (2-μm pitch; see the inset SEM image adapted from Refs. 29 and 30) that consists of parallel lines on the NIST RM8820 artifact. (**C**) Optical image corresponding to another pattern on the NIST artifact with a 4-μm gap that can be barely observed from the brightfield image. (**D**) SEM and darkfield images of a nanoparticle. One cannot find the information of the nanoparticle from the darkfield image because of the low performance of the imaging system.

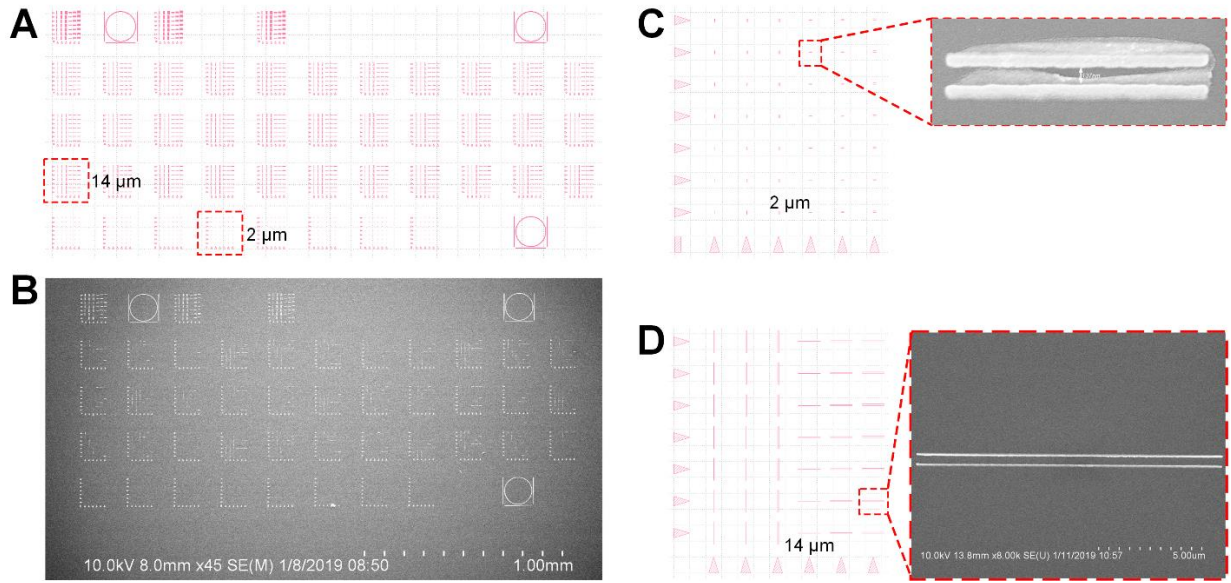

Figure S10: The GDS file and SEM images of the fabricated samples. (**A**) GDS file of the designed nanopatterns. (**B**) A full-view SEM image of the entire set of patterns on the wafer. (**C**) A representative pattern corresponding to the die marked by "2 µm" in **A**. A SEM image corresponding to a double-nanowire structure in **C** shows the fabrication errors induced by the variations in the process. (**D**) A representative pattern corresponding to the die marked by "14 µm" in **A**. The SEM image in **D** shows a typical double-nanowire structure with a nominal length of 14 µm.

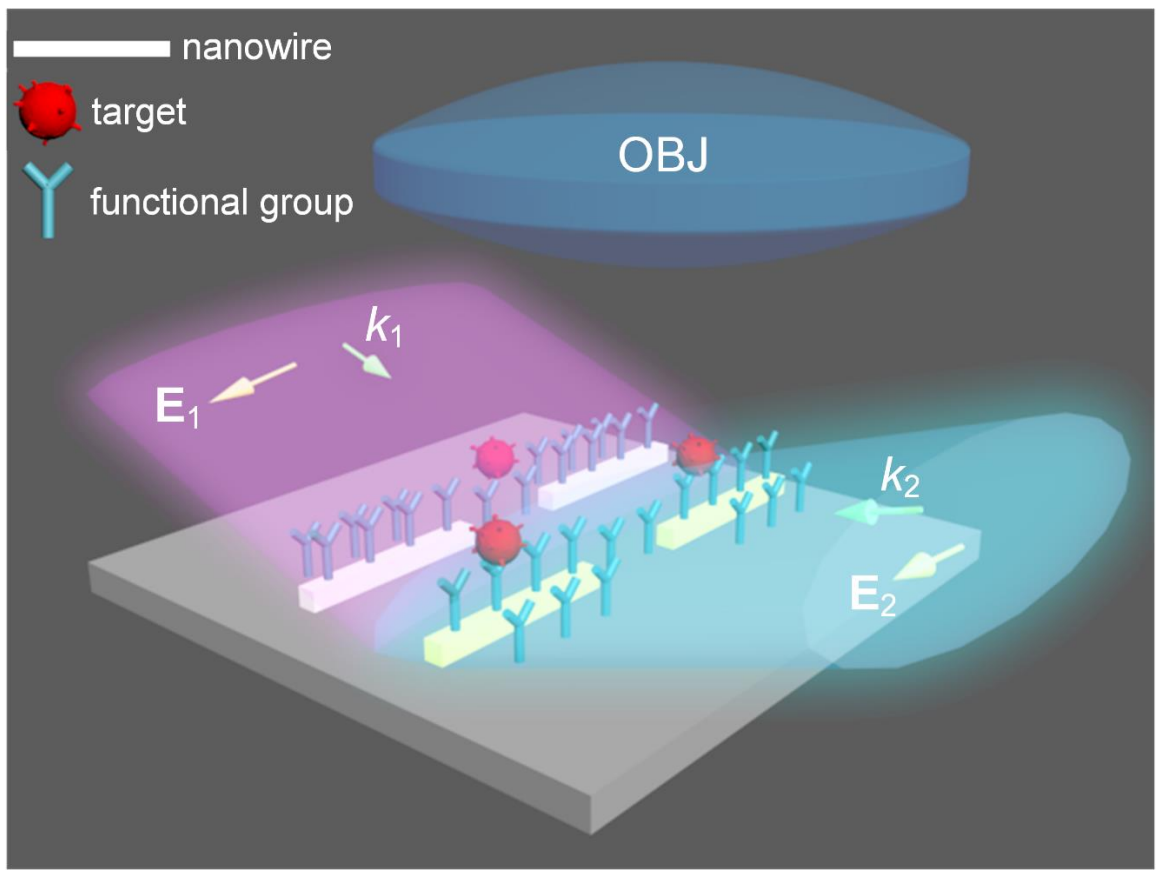

Figure S11: The artwork showing the envisioned sensing of biomaterials using the proposed detection system.

**TABLE I**. **Symmetry of the near-field electric field components for different dipole excitations with respect to the planes $x = 0$ and $y = 0$, respectively.**

| Excitation | $E_x$ | $E_y$ | $E_z$ |
|---|---|---|---|
| Conventional *x*-polarized | **S, S** | A, A | A, S |
| Conventional *y*-polarized | A, A | **S, S** | S, A |
| Conventional *z*-polarized | A, S | S, A | **S, S** |
| Anisotropic *x*-polarized | **A, S** | S, A | S, S |
| Anisotropic *y*-polarized | S, A | **A, S** | A, A |
| Anisotropic *z*-polarized | S, S | A, A | **A, S** |

The dominant field component is in bold. (S: symmetric state; A: anti-symmetric state)